\documentclass[%
 reprint,
superscriptaddress,
 amsmath,amssymb,
 aps,
pra,
hidelinks,
]{revtex4-2}

\usepackage{graphicx}
\usepackage{dcolumn}
\usepackage{textcomp}
\usepackage{bm}
\usepackage{soul,xcolor} 
\usepackage{subcaption}

\usepackage{ragged2e} 
\DeclareCaptionJustification{justified}{\justifying}

\usepackage{orcidlink}

\begin{document}
\sethlcolor{yellow}

\preprint{APS/123-QED}

\title{Metasurface Enhanced Spatial Mode Decomposition}

\author{Aaron W. Jones \orcidlink{0000-0002-0395-0680}}
\email{aaron.jones@ligo.org}
\thanks{A.J, M.W and X.Z contributed equally to this work}
\affiliation{Institute for Gravitational Wave Astronomy, School of Physics and Astronomy, University of Birmingham, Edgbaston, Birmingham B15 2TT, United Kingdom}
\affiliation{OzGrav, University of Western Australia, Crawley, Western Australia, Australia}

\author{Mengyao Wang}%
\email{mengyao.wang@bnu.edu.cn}
\affiliation{Department of Astronomy, Beijing Normal University, Beijing 100875, China}

\author{Xuecai Zhang \orcidlink{0000-0002-7563-3860}}%
\affiliation{Department of Materials Science and Engineering, Southern University of Science and Technology, Shenzhen 518055, China}

\author{\\Samuel J. Cooper \orcidlink{0000-0001-8114-3596}}
\affiliation{Institute for Gravitational Wave Astronomy, School of Physics and Astronomy, University of Birmingham, Edgbaston, Birmingham B15 2TT, United Kingdom}

\author{Shumei Chen \orcidlink{0000-0001-9904-0173}}
\affiliation{School of Science, Harbin Institute of Technology, Shenzhen 518055, China}

\author{Conor M. Mow-Lowry \orcidlink{0000-0002-0351-4555}}
\affiliation{Institute for Gravitational Wave Astronomy, School of Physics and Astronomy, University of Birmingham, Edgbaston, Birmingham B15 2TT, United Kingdom}
\affiliation{Department of Physics and Astronomy, VU Amsterdam, De Boelelaan 1081, 1081, HV, Amsterdam, The Netherlands}
\affiliation{Nikhef, Science Park 105, 1098, XG Amsterdam, The Netherlands}

\author{Andreas Freise \orcidlink{0000-0001-6586-9901}}
\affiliation{Institute for Gravitational Wave Astronomy, School of Physics and Astronomy, University of Birmingham, Edgbaston, Birmingham B15 2TT, United Kingdom}
\affiliation{Department of Physics and Astronomy, VU Amsterdam, De Boelelaan 1081, 1081, HV, Amsterdam, The Netherlands}
\affiliation{Nikhef, Science Park 105, 1098, XG Amsterdam, The Netherlands}

\date{\today}

\begin{abstract}
Acquiring precise information about the mode content of a laser is critical for multiplexed optical communications, optical imaging with active wave-front control, and quantum-limited interferometric measurements. Hologram-based mode decomposition devices, such as spatial light modulators, allow a fast, direct measurement of the mode content, but they have limited precision due to cross-coupling between modes. Here we report the first proof-of-principle demonstration of mode decomposition with a metasurface, resulting in significantly enhanced precision. A mode-weight fluctuation of $6\times 10^{-7}$ was be measured with 1 second of averaging at a Fourier frequency of 80 Hz, an improvement of more than three orders of magnitude compared to the state-of-the-art spatial light modulator decomposition. The improvement is attributable to the reduction in cross-coupling enabled by the exceptional small pixel size of the metasurface. We show a systematic study of the limiting sources of noise, and we show that there is a promising path towards complete mode decomposition with similar precision.
\end{abstract}


\maketitle
A monochromatic laser field is uniquely described by its power and the transverse profile. A standard approach to describe the transverse profile is to decompose it into an orthogonal mode basis, such as the Hermite–Gaussian (HG) or Laguerre–Gaussian (LG) basis. The mode content, the fractional weight and relative phase of different modes, completely describe the laser wave-fronts. Precise determination and manipulation of the mode content can lead to a vast range of new developments in adaptive optics \cite{Pastrana2011,Davies12}, quantum optics \cite{Wagner541} and quantum communications \cite{Wang18,Richardson13}. Even applications that require extremely stable wavefronts, such as laser interferometric gravitational-wave detection will also benefit from accurate information about the mode content \cite{Adhikari12,Brooks16}.

One direct and robust approach to determine the mode content is using correlation filters based on Fourier optics \cite{Golub82,Schulze12}. A particular mode pattern is encoded onto a diffractive optical element and a Fourier imaging system convolves this pattern with an incoming beam. In the far field, the mode weight is proportional to the on-axis intensity \cite{Golub82}. For an orthogonal mode basis, several patterns can be spatially multiplexed by adding a blazed grating to the spatial carrier, allowing the simultaneous interrogation of multiple modes. In practice, the sensors that read the modal powers must have a finite aperture, not only measuring the ideal on-axis intensity \cite{Jones20}. The undesired measurement of some off-axis intensity induces an error, which in this paper, we refer to as the cross-coupling. 

Previous work has demonstrated the success of using spatial light modulators for mode decomposition \mbox{\cite{Schulze12,Forbes16}} and showed an accuracy of mode weighting around 1\,\% \mbox{\cite{Jollivet14}}. This work found $S^2$ imaging to be $\sim 3$ times more precise than correlation filters. Recent work shows that the correlation filter approach can be improved by a factor $\sim 5$ by reducing a using a pinhole to reduce cross-coupling \mbox{\cite{Jones20,PhD.Jones20}}.

The cross coupling effect has been analysed and is limited by the beam-size at the sensor \cite{Jones20,PhD.Jones20}. More precisely,
it scales as \cite{Jones20},
\begin{align}
    \rho_\text{min} \approx \frac{r_a^2}{4w_S^2} + \mathcal{O}\left(\frac{r_a^4}{w_S^4}\right),
\end{align}
where $r_a$ is the radius of the sensing aperture and $w_S$ is the Gaussian beam radius on the sensor. Henceforth, to ultimately suppress the cross coupling, larger beams on the sensor are preferred. As a fact of a Fourier imaging systems \cite{Goodman04} this implies a small incoming beam interrogating at the diffractive optical element, which exhibits a challenging requirement. For example, spatial light modulators have fundamental limit of the pixel resolution. 

\begin{figure}
  \includegraphics[width=0.6\linewidth]{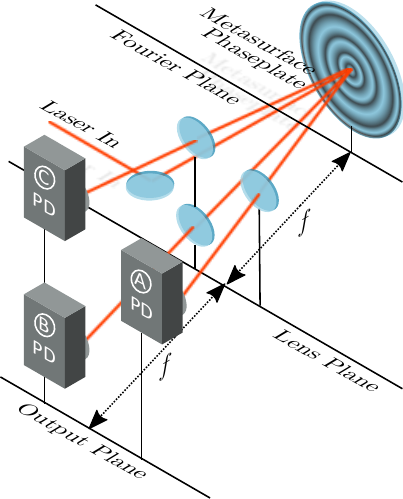}
  \caption{Mode Decomposition Apparatus. The incoming laser strikes the meta-surface and is split into three beams, A, B and C. The lenses focus this light to a large waist in the output plane. Preparatory optics are shown in \ref{fig:bench_layout}.}
  \label{fig:3D}
\end{figure}
Recent developments in the field of optical metasurfaces have led to a novel platform for controlling the multiple degrees of freedom of light at subwavelength scale \cite{Ref1}. The metasurface is an ultrathin structured medium composed of spatially variant plasmonic or dielectric meta-atoms, which can be fabricated by using the state-of-the-art nanofabrication technologies. The electromagnetic response of each meta-atom can be engineered individually, which enables the precise and arbitrary manipulation of the polarization, phase and amplitude of light. In the past decade, the metasurface, as a new kind of planar and multi-functional diffractive optical elements have been extensively explored in the field of planar metalens \cite{Ref2,Ref3}, high efficiency optical holography \cite{Ref4,Ref5}, vortex beam generations \cite{Ref6,Ref7}, quantum information processing \cite{Ref8,Ref9,Ref10,Ref11,Ref12}. One recent proposal uses highly transmissive dielectric metasurface for spatial mode multiplexing on optical fibers (\cite{Kruk2018} and references therein).

In our work, we propose to use a metasurface as the diffractive optical element for mode decomposition. We experimentally addressed the cross-coupling effect and noise issues by encoding three correlation filters for HG00, HG01, and HG10 on one metasurface chip, which allows for a simple calibration procedure and can be easily extended to include more spatial modes. These modes were chosen, as they permit the use Quadrant PhotoDiodes (QPDs) as witness sensors. With the exceptionally small pixel size of the metasurface, we are able to measure the first-order mode content with a noise floor $< 10^{-6} / \sqrt{\text{Hz}}$ above of 25\,Hz. 
\begin{figure}
\centering
\includegraphics[width=\linewidth]{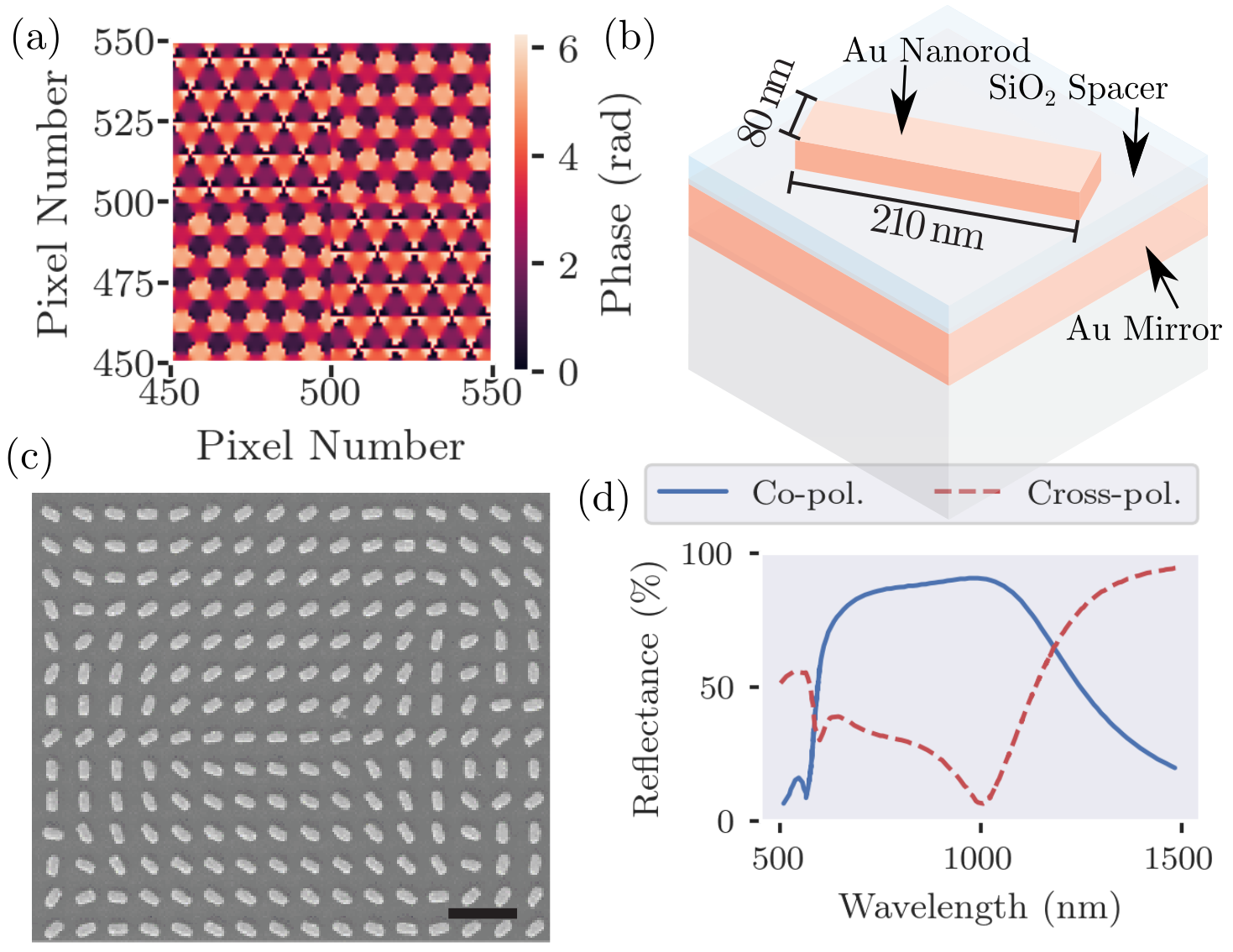}
  \caption{Design and Fabrication of the Au/SiO\textsubscript{2}/Au tri-layer metasurface. (a) The central 100 pixels of the phase distribution encoded on the MODAN metasurface. (b) The illustration of the tri-layer meta-atom structure. (c) The scanning electron microscope (SEM) image of the metasurface. Scale bar: 720 nm. (d) The simulated Co-polarisation and Cross-polarization in the reflected beam. The incident light is LCP.}
  \label{fig:meta}
\end{figure}

Figure~\ref{fig:3D} shows an overview of the mode decomposition apparatus (MODAN). Phase maps of HG00, HG01, and HG10 together with three different blazed gratings are encoded on the metasurface chip. The entire metasurface chip active area was 500 \textmu m $\times$ 500 \textmu m with $10^3 \times 10^3$ pixels for 1064\,nm laser beam. The incoming laser, strikes the metasurface and is split into three beams, A, B and C.  The on-axis intensity of beam A corresponds to the power in the HG00 mode, while B and C correspond to HG01 and HG10 respectively. A lens focuses these beams to a large waist where the on-axis power is then measured using inexpensive standard small aperture photo-diodes. This photodiode had a circular 250\,\textmu m aperture and a quantum efficiency of 70\% \cite{PDA05CF2}. Cross coupling could have been further suppressed by using a smaller aperture prior to the photodiode. A 10\,\textmu m aperture was tested but the dark noise of the photodiode was in excess of any measurable signal. The signal was digitized in real time using a state-of-the-art, low noise multi-channel digitizer \cite{cds}. This system is a clone of the digitizer used by LIGO. The combination of these technologies ensured that digital quantization noise was not limiting, while also permitting a measurement bandwidth from 1\,mHz to 10\,kHz.    

Figure~\ref{fig:meta} shows the phase distribution of the MODAN metasurface. The phase map was computed by adding the phase of the Hermite-Gauss mode profiles to blazed gratings, each rotated by $2\pi/3$. For mathematical details, see Supplementary \ref{sec:phasemap}. To achieve the required phase distribution, the meta-atom is designed based on the concept of geometric Pancharatnam-Berry phase \cite{Ref13}. As shown in Figure~\ref{fig:meta}, the meta-atom has a Au nanorod/SiO\textsubscript{2} spacer/Au mirror tri-layer configuration. The width, length and height of the gold nanorod is 80 nm, 210 nm and 30 nm, respectively. The thickness of the SiO\textsubscript{2} spacer and gold reflector layer are 90 nm and 150 nm, respectively. Under the incidence of left or right-circularly polarized (LCP/RCP, $\sigma=\pm 1$) light, the reflected light with opposite handedness (RCP/LCP) can be obtained by choosing the anisotropic meta-atom with appropriate geometrical parameters. By rotating the fast axis of the gold nanorod, the geometric phase experienced by the reflected light with RCP/LCP polarization state can be described by $\phi=2\sigma \theta$, where $\theta$ represents in-plane orientation angle of the gold nanorod. It is found that $\phi$ can be continuously tuned from 0 to 2$\pi$. The period of the meta-atom is 360 nm. The tri-layer metasurface, which is shown by the scanning electron microscope image in Figure~\ref{fig:meta}, was fabricated by using the electron beam lithography and metal lift-off processes. The optical properties of a metasurface unit cell is calculated by using the commercial finite-difference time-domain (FDTD) software (Lumerical FDTD Solutions). As shown in Figure~\ref{fig:meta}, the cross-polarization conversion efficiency at the wavelength of 1064 nm reaches over 80\% under the incidence of left circularly polarized light, which indicates the high optical efficiency of the tri-layer metasurface device.
\begin{figure}
\centering
\begin{subfigure}[b]{0.48\textwidth}
  \centering
  \includegraphics[width=0.8\textwidth]{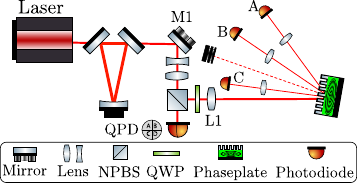}
  \caption{Validation apparatus. Laser light is filtered though a mode cleaner cavity. A small angular modulation is applied at M1. The beam is then incident on the two devices under test: a QPD and the metasurface phaseplate}
  \label{fig:bench_layout}
\end{subfigure}
\hfill
\begin{subfigure}[b]{0.48\textwidth}
  \centering
  \includegraphics[width=\textwidth]{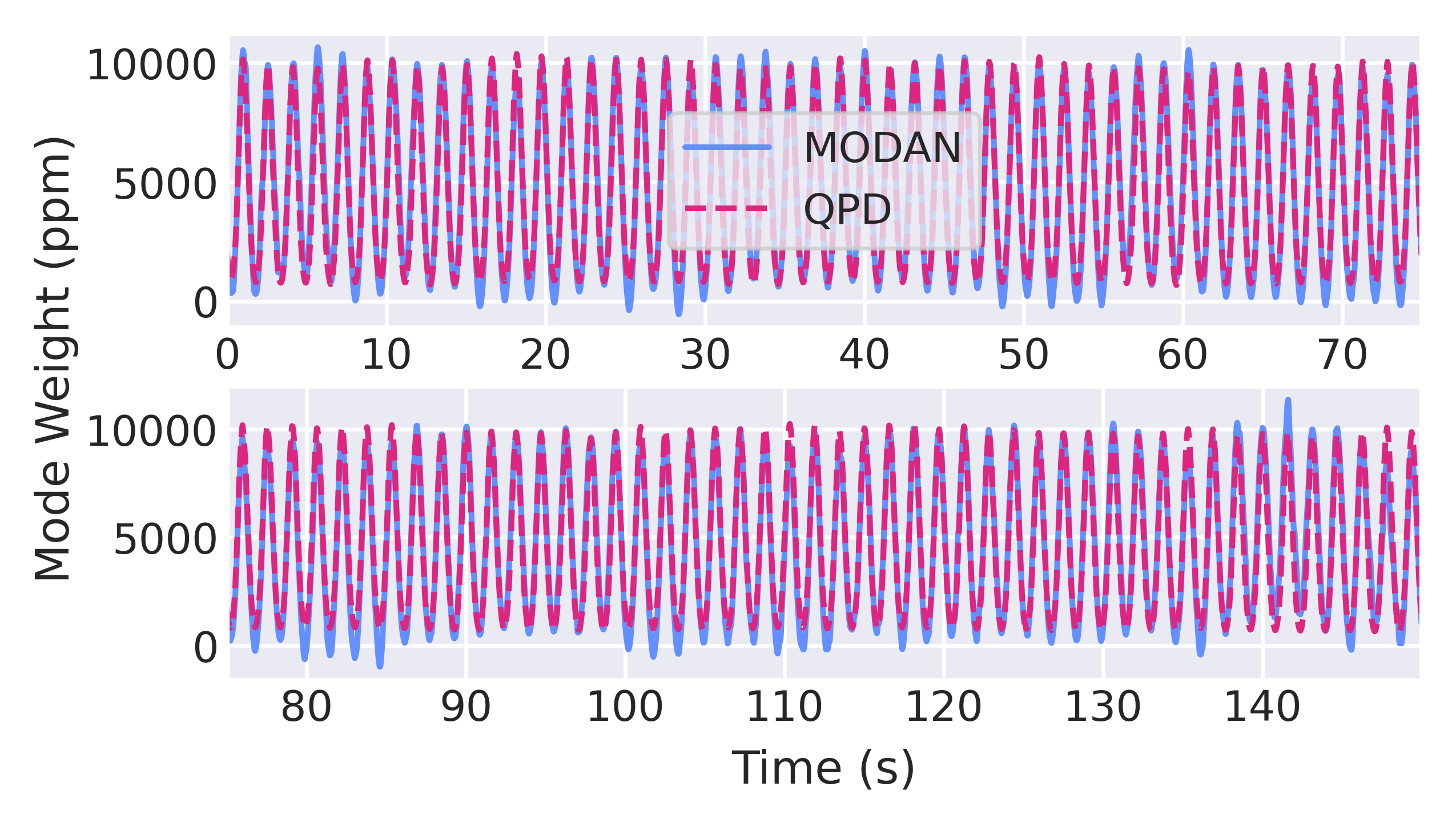}
  \caption{\small Calibration Signal. 150\,s of data was gathered for the calibration procedure. This data shows excellent coherence between the QPD and MODAN based sensors at the 318\,mHz modulation frequency.}
  \label{fig:cohere}
\end{subfigure}
\caption{Calibration and Validation}
\label{fig:c_and_v}
\end{figure}
To validate our sensor, we first prepared a clean beam by spatially, polarization and spectrally filtering laser light though a rigid triangular mode cleaner cavity \cite{WUGBKSS98,UeharaSPIE} as illustrated in Figure \ref{fig:bench_layout}. The metasurface and photodiodes were calibrated by injecting a small amount of HG10 at 318 mHz, realized by an angular modulation of steering mirror \textit{M1}. This technique works as a shifted HG00 can be described as an addition of HG10 mode. The modulated light was split with one branch going to a witness QPD, which required a 1.2\,mm waist radius beam. The second branch was directed towards the metasurface, which required a 55\,\textmu m waist radius of elliptically polarized light. {The amount of HG10 does not change on propagation, however, the QPD is only sensitive the the real component of the HG10. Therefore, Gouy phase information was required to convert the QPD signal into mode weight. Additionally, the origin of the co-ordinate system for the QPD and the metasurface did not coincide at the part-per-million (ppm) level. Since the QPD measures a mode amplitude rather the power, this could be accounted by a small offset during signal processing. MODAN channels A and C were used to compute the MODAN mode weight. Due to exceptional metasurface resolution in nano-scale size, cross coupling was substantially suppressed. However, at the ppm mode weighting which we present in this work, a residual leakage still occurred from the TEM00 mode. As a result, we subtracted some of the DC leakage from our results during analogue signal processing, the remaining offset was computed and subtracted in digital post processing. See the Appendix~\ref{sec:calib} for more details on the calibration. After calibration, we recorded 150 seconds of data respectively and compared their performances.} We found the coherence between the two devices was good, as shown in Figure \ref{fig:cohere}. However, noise in the power supply caused the offset to drift slightly during the measurement, leading to increased measurement uncertainty at low frequencies.

\begin{figure}
\centering
\includegraphics[width=0.5\textwidth]{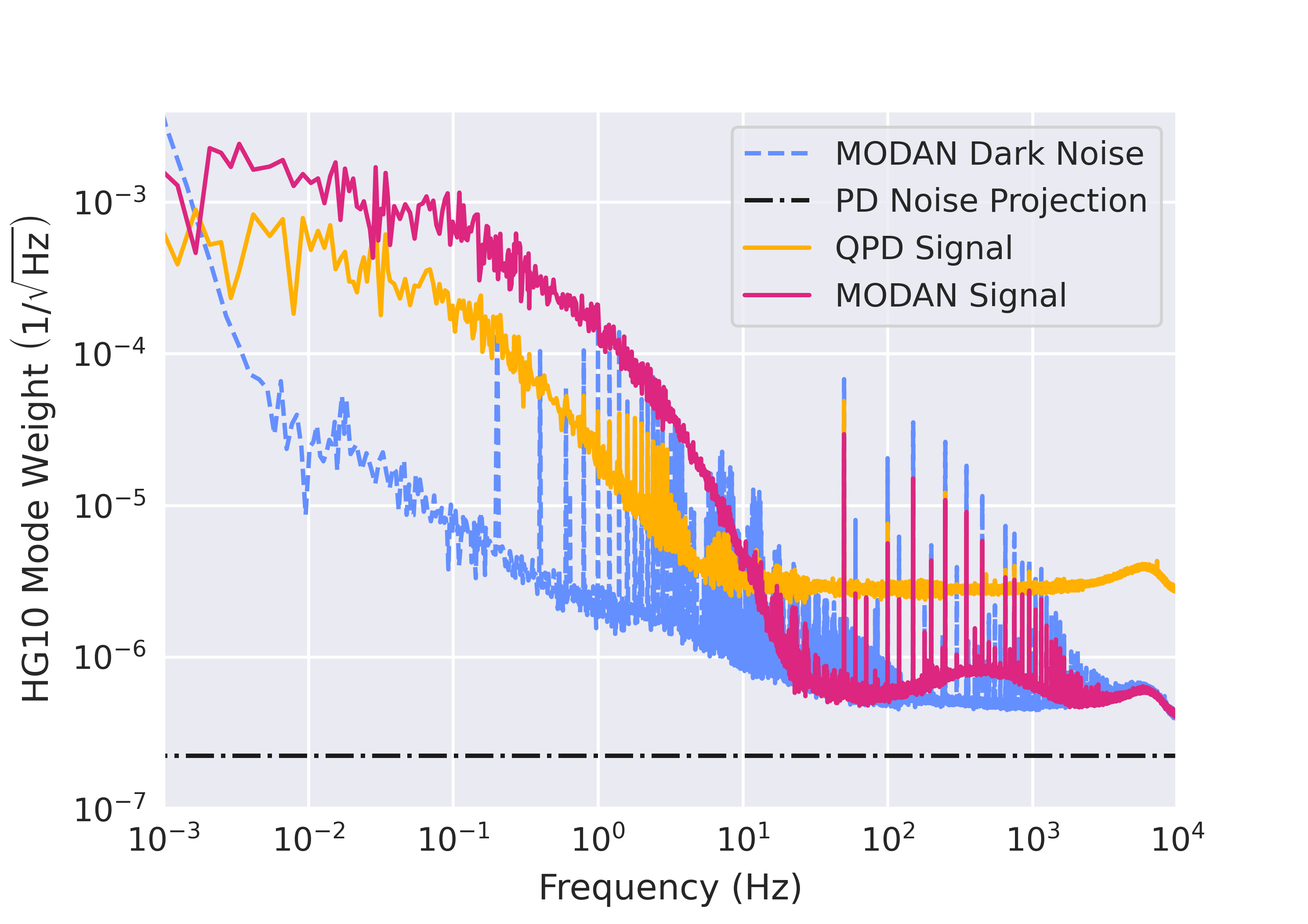}
\caption{\small Noise contributions. \textit{MODAN Signal} and \textit{QPD Signal} show the measured signals, whilst the beam is centered on the MODAN, but out of any alignment control loop. The manufacturers photo-diode noise estimate from Channels A and C is shown by \textit{PD Noise Projection}. \textit{Dark Noise} shows the total noise of the MODAN photo-diodes, signal processing and digitizer electronics, with no laser illumination.}
\label{fig:asd}
\end{figure}

One powerful way of understanding these noises is amplitude spectral density (ASD) plot, shown in Figure~\ref{fig:asd}. This shows the noise in the HG10 mode weight measurement by the MODAN and the QPD respectively, as a function of frequency. Thus, indicating what minimum mode weight is measurable with an appropriate bank of electronic filters. The noise on an arbitrary duration measurement is given by the integral of this curve, over the frequencies of interest. \textit{MODAN Signal} and \textit{QPD signal} were measured simultaneously using the calibration described for Figure \ref{fig:cohere}. With no laser illumination, the \textit{Dark Noise} gives the total noise floor of the MODAN photo-diodes, signal processing and digitizer electronics. 

At high frequency, we see that the metasurface based sensor outperforms the QPD, achieving the sensitivity of $6\times10^{-7}/\sqrt{\mathrm{Hz}}$ at 80\,Hz. Another way of writing this is that the mode weight fluctuation at 80\,Hz can be measured down to 0.6\,ppm with signal-to-noise ratio 1 and 1 s averaging. At these high frequencies, the measurement is limited by a white electronic noise. We expect that the photodiodes self-noise (shown in orange) was slightly in excess of that estimated by the manufacturer, which explains this limitation.  The feature at 6 kHz in all three data traces is an artifact of the digitizer anti-aliasing filter. 

At lower frequencies, the measurement is limited by an optical effect. We expect that this noise is residual beam drift, possibly caused by seismic, thermal and air pressure fluctuations causing an apparent motion of the lens, with respect to the beam axis defined by the metasurface and aperture. This effect shifts the centre of beam away from the aperture, thus increasing cross-coupling. This means that over long timescales ($t>100$\,s) the RMS error is $<$500\,ppm with no averaging, adding 100\,s of averaging reduces this to $<200$\,ppm.

\section*{Conclusions}
In this work, we demonstrate a metasurface enhancement to correlation filter mode decomposition. This enhancement permits the investigation of small mode weights among a larger carrier mode, in this case TEM00. This is particularly useful in precision metrology, where high levels of mode matching are required. Furthermore, sub-micron pixels enable a reduction in unused diffraction orders, improving power efficiency, thus reducing shot noise and increasing the available spatial multiplexing. Therefore the signal to dark-noise ratio is increased. In contrast to QPD’s and Bulls-Eye photodetectors, correlation filters can be trivially engineered for the measurement of an arbitrary mode pattern \cite{Forbes16}. However, due to power conservation, the simultaneous interrogation of multiple modes reduces the power on each branch, thus decreasing the signal to noise ratio.

Applications requiring high-frequency mode-decomposition, such as mode-division-multiplexing and characterization of single-mode fibers, may reduce cross-talk using the metasurface enhancement. These systems are likely to be limited by electronic noises in the photo-diode. An improved photo-diode may permit shot noise limited sensitivity.

Applications requiring low frequency mode analysis, such as correction of thermally induced mode mismatches in high power systems, will need to carefully consider the low frequency the stability of the phase-plate, lens, photo-diode and electronics to achieve the very highest dynamic ranges. Gravitational wave detectors may implement this technique to monitor parametric instabilities.

Future work may wish to combine the advantages of metasurfaces with adaptive optics, in the context of mode decomposition.  

\begin{acknowledgments}
\paragraph{Thanks} The authors jointly thank John Bryant and David Hoyland 
 for developing the low noise QPD and modifications to the
 EUCLID digitizer, used in precursor experiments. The authors
 thank Dr Artemiy Dmitriev for developing CDS units at Birmingham.
 A. W. Jones was supported by an EPSRC studentship and ARC Centre of Excellence for Gravitational Wave Discovery (OzGrav), project number CE170100004.
 
\paragraph{Author Contributions} A.J., A.F. and M.W. designed the study and
 developed the phase-pattern. X.Z. and S.Ch. developed and fabricated the
 metasurface surface. A.J. design, assembly and operation of optical testing apparatus.
 A.J. and C.M. designed the analogue signal processing and
 data analysis. S.Co. developed, built and operated the digital acquisition. 
 A.J., C.M. and M.W. drafted the manuscript.
\end{acknowledgments}

\appendix

\section{Determination of Phase Pattern}
\label{sec:phasemap}
The phase map was computed from the Hermite-Gauss amplitude functions, $u(x,y,z)_{n,m}$ and blazed gratings. The equation for the blazed grating is, 
\begin{align}
    B(x,y)_{n,m} = \frac{2\pi}{\Lambda}(\cos(\gamma_{n,m})x + \sin(\gamma_{n,m})y,
\end{align}
where $\gamma_{n,m}$ is the rotation angle for this branch and $\Lambda$ is the grating period. The complete phase map is then given by
\begin{align}
    P_c(x,y) &= \pi \\\nonumber
    &+ \arg\left[\sum_{n,m} \exp(i \arg(u_{n,m}(x,y)|_{z_0}) + B(x,y)\right],
\end{align}
where the mode function is evaluated at the waist position, $z=z_0$. In this work, $\gamma_{0,0}=0$ $\gamma_{0,1}=\frac{2\pi}{3}$, $\gamma_{0,1}=\frac{4\pi}{3}$. 

\section{Calibration Procedure for Meta-Phase plate}
\label{sec:calib}
The measurement of modal weight presented in the main text, uses a photo-diode to determine the amount of optical power in a small region close to the axis of propagation. This photo-diode outputs a current, which is proportional to the optical power illuminating the diode. This current must be amplified, converted into a voltage, filtered, digitized and analyzed spectrally to produce the data we present. These steps must be carefully understood to convert the resulting digital data back into a meaningful mode weight. We refer to this process as sensor calibration.

A detailed calibration procedure, from first principals, with a different modulation frequency (20\,Hz) and different digitizer, can be found in \cite{PhD.Jones20}. In this, Jones demonstrates that the calibration obtained from first principals and the calibration against the witness QPD are consistent. However, during such a calibration errors on photo-diode aperture and electrical gains dominate the uncertainty on the calibration. For this reason, we opted to calibrate against the witness sensor. 

Contrary to the experimental description in \cite{PhD.Jones20}, for this work we used a clone of the LIGO CDS system \cite{cds} to simultaneously record several data channels from the experiment. With this system, all data was gathered at the same clock rate, simplifying the analysis and allowing witness channels for diagnostics. The CDS system itself had been pre-calibrated to output four voltages of interest: the difference in the left and right halves of the QPD, $V_{dx}$, the total power on the QPD, $V_{SUM}$, and the voltage output by the MODAN photo-diodes, $V_{10}$ and $V_{00}$.

This data was taken in three conditions: \textit{modulated}, where the alignment of the input beam was modulated at 318\,mHz, denoted by $^{M}$; \textit{unmodulated}, where the beam was still aside from beam jitter, denoted by $^{UM}$; and \textit{dark} where the laser was switched off, but all other electronics was active, denoted by $^{D}$. The dark data was used to compute electrical offsets and noise levels. The modulated QPD data was used to produce a calibration factor for the MODAN. Finally, we use this calibration factor to produce the unmodulated MODAN data.

\subsection{Removal of Offsets}
The photo-diode and amplification stages contain some small voltage offsets. These offsets could trivially be calculated for the HG10 and HG00 channels from the average value of the dark noise,
$$
O_{PD} = \frac{\sum_{i=0}^{N^D} V^{D}_{PD}}{N^D},
$$
where $N^D$ is the number of elements in the dark data trace and $PD = \{01,00\}$.

\subsection{Mode Power to Mode Weight}
The MODAN photo-diodes produce a current which is proportional to mode power. In this case, we remove power fluctuations and present a mode weight. Since the modulation is small, we can approximate that the power on the HG00 sensor depends only on laser power. Therefore, the mode HG10 mode weight measured by the MODAN is,
$$
\rho_{10}^{MD} = C_\text{calib.}\frac{V_{10} - O_{10}}{V_{00} - O_{00}},
$$
where $C_\text{calib.}$ is the calibration factor we need to compute.

\subsection{Calibration}
As shown in \S4.3.2 of \cite{PhD.Jones20}, the HG10 mode amplitude is given by,
$$
a_{10}^{QPD, M} = \frac{V^{M}_{10}}{V^{M}_{SUM}} \frac{\pi}{2\sqrt{2}\cos(40 \mathrm{deg})},
$$
where 40 degrees is the Gouy phase difference between the point of angular actuation and the QPD. Comparing the above two equations, $C_\text{calib.}$ is then,
$$ 
\frac{1}{C_\text{calib.}}\left(a_{10}^{QPD, M} - \Delta a \right)^2 \xrightarrow[\mathrm{fit}]{} \frac{V_{10}^M - O_{10}}{V_{00}^M - O_{00}}
$$
where $\Delta a$ is a DC alignment offset between the sensors in units of mode amplitude.

%

\end{document}